\documentstyle[12pt]{article}
\begin{document}

\centerline{\Large\bf Magnetization in 2+1 dimensional QED at }
\centerline{\Large\bf Finite Temperature and Density}

\vskip 10mm
\centerline{Jens O. Andersen and Tor Haugset}
\centerline{\it Institute of Physics}
\centerline{\it University of Oslo}
\centerline{\it P.O. BOX 1048, Blindern}
\centerline{\it N-0316 Oslo, Norway}

\vspace{10mm}
{\footnotesize We consider Dirac fermions moving in a plane with a static
homogeneous magnetic field orthogonal to the plane. We calculate the
effective action at  finite temperature and density. The magnetization is
derived and it is shown that the fermion gas exhibits de Haas-van Alphen
oscillations at small temperatures and weak magnetic fields.
We also comment upon earlier work. }

\section{Introduction}
Many of the phenomena that have been discovered in condensed matter
physics
over the last few decades are to a very good approximation two
dimensional.
The most important of these are the (Fractional) Quantum Hall effect and
high $T_{c}$
superconductivity \cite{qhe}.

Quantum field theories in lower dimensions have therefore become of
increasing
interest in recent years. Both systems mentioned above have been
modelled by anyons, which are particles or excitations that obey
fractional
statistics. Anyons can be described in terms of Chern-Simons field
theories
[1-3]. The literature on 2+1d field theory in general and Chern-Simons
field
theory in particular is now vast (see e.g
refs. [1-5]),
and we shall only comment upon a few selected papers, which are relevant
for the present work.

Ten years ago Redlich \cite{reddik} considered fermions in a plane moving
in a constant electromagnetic field. Using Schwinger's proper time
method \cite{schwing}
to obtain the effective action for the gauge field, he demonstrated that
a Chern-Simons
term was induced by radiative corrections. The Chern-Simons term is parity
breaking and is gauge-invariant modulo surface terms.
More recently, Lykken {\em et al.}
\cite{lykken} have examined two dimensional Dirac fermions coupled to a
gauge field
whose dynamics is governed by a (topological) Chern-Simons term. It was
found
that this planar Fermi gas becomes superconducting (below a critical
temperature) when the induced
Chern-Simons term exactly cancels the one appearing in the classical
action.

External electromagnetic fields may give rise to induced charges in the
Dirac vacuum if the energy spectrum is asymmetric with respect to some
arbitrarily chosen zero point. The vacuum charge comes
about since the number of particles gets reduced (or increased) relative
to
the free case. Furthermore, induced
currents may appear and are attributed to the drift of the induced
charges.
This only happens if the external
field does not respect the translational symmetry of the system.
These interesting
phenomena have been examined  in detail by Flekk{\o}y and Leinaas
\cite{lein}
in connection with
magnetic vortices and their relevance to the Hall effect has been studied
by Fumita and Shizuya \cite{japs}.

In the present paper we re-examine the system considered by
Redlich \cite{reddik}. We shall restrict ourselves to the
case of a constant magnetic field, but we extend the analysis by including
thermal effects and we shall mainly focus on the magnetization of the
system.

In the first section we find the
solutions of the Dirac equation for particles in a constant magnetic
field.
These solutions are then used to calculate the
thermal fermion propagator, from which we derive
the effective action for the gauge field.
In the fourth section we compute the magnetization. It is shown that the
de Haas-van Alphen oscillations  [10,11]
occur at low temperatures and weak magnetic fields, and that is the main
result of this paper.
We consider various limits of the magnetization
that may be understood from a physical point of view and we explicitly
demonstrate that we get the correct high temperature limit
(i.e. free fermions).

Finally, we
summarize and draw some conclusions in the last section.
\section{Dirac Equation}
In this section we shall briefly discuss the solutions of the Dirac
equation
in 2+1d with a constant magnetic field along the $z$-axis. The Dirac
equation reads
\begin{equation}
\label{eq:dirac}
(i\gamma^{\mu}\partial_{\mu}-e\gamma^{\mu}A_{\mu}-m)\psi=0,
\end{equation}
where the gamma matrices satisfy the Clifford algebra
\begin{equation}
\label{eq:clif}
\{ \gamma^{\mu},\gamma^{\nu}\}=2g^{\mu\nu}.
\end{equation}
In 2+1 dimensions the fundamental representation of the Clifford algebra
is
given by $2\times 2$ matrices and these can be constructed from the
Pauli matrices. Furthermore, in 2+1d there are two inequivivalent
choices of
the gamma matrices, which corresponds to
$\gamma^{\mu}\rightarrow -\gamma^{\mu}$. From eq. (\ref{eq:dirac}) we see
that this extra degree of freedom may be absorbed in the sign of $m$.
These choices correspond to ``spin up'' and ``spin down'', respectively.

We would also like to point out that the angular momentum operator
{\bf $\sigma$} is a  pseudo vector
in 2+1d, implying that the Dirac equation written in terms of these
matrices
does not respect parity. This is no longer the case if the Dirac equation
is expressed in terms of $4\times 4$ matrices. The latter
representation is
reducible and reduces to the two inequivalent fundamental representations
mentioned above. Let us explicitly verify this. A natural choice is the
three block diagonal $4\times 4$ matrices
\begin{equation}\gamma^{0}_{4}=
\left(\begin{array}{cc}
\gamma^{0}&0\\
0&-\gamma^{0}\\
\end{array}\right),\hspace{1cm}\gamma^{1}_{4}=
\left(\begin{array}{cc}
\gamma^{1}&0\\
0&-\gamma^{1}\\
\end{array}\right),\hspace{1cm}\gamma^{2}_{4}=
\left(\begin{array}{cc}
\gamma^{2}&0\\
0&-\gamma^{2}\\
\end{array}\right),
\end{equation}
where $\gamma^{0}$, $\gamma^{1}$ and $\gamma^{2}$ themselves are
$2\times 2$ matrices satisfying
the
algebra.
This representation trivially reduces to the sum of the two inequivalent
choices of the $2\times 2$ gamma matrices. Furthermore, the Dirac equation
written in terms of the above matrices respects parity (a parity
transformation simply interchanges the coupled equations and thereby the
components of the spinor).

In the following we make the choice $\gamma^{0}=\sigma_{3}$,
$\gamma^{1}=-i\sigma_{2}$, $\gamma^{2}=-i\sigma_{1}$ and  we have chosen
the asymmetric gauge $A_{\mu}=(0,0,-Bx)$. The metric is
diag\,($-1$,\,1,\,1).

The Dirac equation then takes the form
\begin{equation}
\label{eq:d2}
\left(\begin{array}{cc}
i\frac{\partial}{\partial t}-m&i\frac{\partial}{\partial x}-ieBx+
\frac{\partial}{\partial y} \\
-i\frac{\partial}{\partial x}-ieBx+\frac{\partial}{\partial y} &
-i\frac{\partial}{\partial t}-m\\
\end{array}\right)\psi ({\bf x},t) =0.
\end{equation}
Here $\psi ({\bf x},t)$ is a two component spinor.
We now assume that the wavefunctions may be written as
\begin{equation}\psi_{\kappa}({\bf x},t)=\exp (-iEt+iky)
\left(\begin{array}{c}
f_{\kappa}(x) \\
g_{\kappa}(x)  \\
\end{array}\right),\hspace{1cm}
\end{equation}
where ${\kappa}$ denotes all quantum numbers necessary in order to
completely
characterize the solutions .
Inserting this into eq. (\ref{eq:d2}) one obtains
\begin{equation}
\label{eq:upd}
\left(\begin{array}{cc}
E-m&-\xi_{+} \\
\xi_{-}&-E-m\\
\end{array}\right)=
\left(\begin{array}{c}
f_{\kappa}(x) \\
g_{\kappa}(x) \\
\end{array}\right)
\end{equation}
where
\begin{equation}
\label{eq:xi}
\xi_{\pm}=-i\partial_{x}\mp i(k-eBx).
\end{equation}
The equation for $f_{\kappa}(x)$ is readily found from eq. (\ref{eq:upd}):
\begin{equation}
\label{eq:fx}
\left ( E^{2}-m^{2} -\xi_{+}\xi_{-}\right ) f_{\kappa}(x)=0.
\end{equation}
The eigenfunctions of $\xi_{+}\xi_{-}$, provided that $eB>0$, are
\cite{koba}
\begin{equation}
\label{eq:in}
I_{n,k}=(\frac{eB}{\pi})^{\frac{1}{4}}\exp\left [\, -\frac{1}{2}
(x-\frac{k}{eB} )^{2}eB\,\right ]\frac{1}{\sqrt{n!}}H_{n}
\left [\,  \sqrt{2eB} (x-\frac{k}{eB})\,\right ],
\end{equation}
where
$H_{n}(x)$ is the $n$'th Hermite polynomial.
Furthermore, $I_{n,k}(x)$ is normalized to unity and  satisfies
\begin{eqnarray} \nonumber
\xi_{-}I_{n,k}(x)&=&-i\sqrt{2eBn}I_{n-1,k}(x), \\  \nonumber
\xi_{+}I_{n,k}(x)&=&i\sqrt{2eB(n+1)}I_{n+1,k}(x).
\end{eqnarray}
Combining eqs. (\ref{eq:fx}) and (\ref{eq:in}) yields
\begin{equation}
f_{\kappa}(x)=I_{n,k}(x),\hspace{1cm}E^{2}=m^{2}+2eBn.
\end{equation}
The function $g_{\kappa}(x)$ satisfies
\begin{equation}
g_{\kappa}(x)=\frac{\xi_{-}}{E+m}f_{\kappa}(x),
\end{equation}
implying that
\begin{equation}
g_{\kappa}(x)=-i\sqrt{2eBn}I_{n-1,k}(x).
\end{equation}
The normalized eigenfunctions become
\begin{equation}
\label{eq:eigenm}
\psi_{n,k}^{(\pm)} ({\bf x},t)=\exp (\mp iE_{n}t+iky)\sqrt{\frac{E_{n}
\pm m}{2E_{n}}}
\left(\begin{array}{c}
I_{n,k}(x) \\
\frac{\mp i\sqrt{2eBn}}{E_{n}\pm m}I_{n-1,k}(x) \\
\end{array}\right),
\end{equation}
where $n=0,1,2,...$, $E_{n}=\sqrt{m^{2}+2eBn}$ and $\psi_{n,k}^{(\pm)}
({\bf x},t)$
are positive and negative energy solutions, respectively.
Note that $\psi^{(-)}_{0,k}({\bf x},t)=0$ and that we have defined
$I_{-1,k}(x)\equiv 0$.
The spectrum is therefore asymmetric and this asymmetry is intimately
related
to the induced vacuum charge, as will be shown in section 4.
In fig.~\ref{spec} a)
we have shown the spectrum for $m>0$ and  in fig.~\ref{spec} b) for
$m<0$.
\\ \\
The field may now be expanded in the complete set of eigenmodes:
\begin{equation}
\label{eq:expa}
\Psi ({\bf x},t)=\sum_{n=0}^{\infty}\int dk \left [\,b_{n,k}
\psi_{n,k}^{(+)} ({\bf x},t)+ d_{n,k}^{\,\ast}\psi_{n,k}^{(-)}
({\bf x},t)\,\right ].
\end{equation}
Quantization is carried out in the usual way by promoting the Fourier
coefficients to operators satisfying
\begin{equation}
\{b_{n,k},b_{n^{\prime},k^{\prime}}^{\dagger}\}=\delta_{n,n^{\prime}}
\delta_{k,k^{\prime}},\hspace{1cm}\{d_{n,k},d_{n^{\prime},
k^{\prime}}^{\dagger}\}=\delta_{n,n^{\prime}}\delta_{k,k^{\prime}},
\end{equation}
and all other anti-commutators being zero.
\section{Fermion Propagators and the Effective Action}
In the previous section we solved the Dirac equation and with the wave
functions at hand, we can construct the propagator. From the trace of the
propagator the effective action to one-loop order is calculated.\\  \\
{\it The fermion propagator.}$\,\,$
The fermion
propagator in vacuum is
\begin{equation}
iS_{F}(x^{\prime},x)=\langle 0\!\mid T\Big [\,\Psi ({\bf x}^{\prime},
t^{\prime})\overline{\Psi} ({\bf x},t) \,\Big ]\mid\!0\rangle,
\end{equation}
where $T$ denotes time ordering. By use of the expansion (\ref{eq:expa})
one finds
\begin{equation}
iS_{F}(x^{\prime},x)=\sum_{n=0}^{\infty}\int \frac{dk}{2\pi}\left
[\, \theta (t^{\prime}-t)\psi^{(+)}_{n,k}({\bf x}^{\prime},t^{\prime})
\overline{\psi}_{n,k}^{(+)}({\bf x},t)-\theta (t-t^{\prime})
\psi^{(-)}_{n,k}({\bf x}^{\prime},t^{\prime})
\overline{\psi}_{n,k}^{(-)}({\bf x},t)\,\right ].
\end{equation}
After some purely algebraic manipulations and using the integral
representation of the step function, we obtain
\begin{equation}
\label{eq:fp}
S_{F}(x^{\prime},x)_{ab}=-\frac{1}{4\pi^{2}}\sum_{n=0}^{\infty}
\int dkd\omega \frac{E_{n}+m}{2E_{n}}
\exp\left [\,-i\omega (t^{\prime}-t)+ik(y^{\prime}-y)\,\right ]
\frac{1}{\omega^{2}-E^{2}_{n}+i\varepsilon}S_{ab}(n,\omega,k).
\end{equation}
Here $S_{ab}(n,\omega,k)$ is the matrix
\begin{equation}
\label{eq:matrix}
\left(\begin{array}{cc}
I_{n,k}(x^{\prime})I_{n,k}(x)&\frac{E_{n}}{E_{n}+m}
\sqrt{2eBn}I_{n,k}(x^{\prime})I_{n-1,k}(x)\\
\frac{E_{n}}{E_{n}+m}\sqrt{2eBn}I_{n-1,k}(x^{\prime})
I_{n,k}(x)&I_{n-1,k}(x^{\prime})I_{n-1,k}(x)\\
\end{array}\right).
\end{equation}
At finite temperature and chemical potential we write the thermal
propagator as
(see ref. \cite{per} for details)
\begin{equation}
\langle S_{F}(x^{\prime},x)\rangle_{\beta ,\mu}=S_{F}(x^{\prime},x)+
S_{F}^{\beta ,\mu}(x^{\prime},x).
\end{equation}
The thermal part of the propagator is
\begin{equation}
\label{eq:thermpro}
iS_{F}^{\beta,\mu}(x^{\prime},x)=-\sum_{n=0}^{\infty}\int
\frac{dk}{2\pi}\left [\,f^{+}_{F}(E_{n})
\psi^{(+)}_{n,k}({\bf x}^{\prime},t^{\prime})
\overline{\psi}_{n,k}^{(+)}({\bf x},t)-f^{-}_{F}(E_{n})
\psi^{(-)}_{n,k}({\bf x}^{\prime},t^{\prime})
\overline{\psi}_{n,k}^{(-)}({\bf x},t)  \,\right ],
\end{equation}
where
\begin{equation}
f^{(+)}_{F}(\omega)=\frac{1}{\exp \beta(\omega-\mu)+1},
\hspace{1cm}f^{(-)}_{F}(\omega)=
1-f^{(+)}_{F}(-\omega)=\frac{1}{\exp \beta(\omega+\mu)+1}.
\end{equation}
This may be rewritten as
\begin{equation}
S_{F}^{\beta,\mu}(x^{\prime},x)=i\sum_{n=0}^{\infty}\int
\frac{dk}{2\pi}d\omega\,\exp ik(y^{\prime}-y)
\exp i\omega(t^{\prime}-t)f_{F}(\omega)
\delta (\omega^{2}-E^{2}_{n}-i\varepsilon)S_{ab}^{\beta}(n,\omega,k).
\end{equation}
Here $S_{ab}^{\beta}(n,\omega,k)$ is the matrix
\begin{equation}
\left(\begin{array}{cc}
(\omega +m)I_{n,k}(x^{\prime})I_{n,k}(x)&I_{n-1,k}(x^{\prime})
I_{n,k}(x)\\
I_{n,k}(x^{\prime})I_{n-1,k}(x)&(\omega-m)I_{n-1,k}(x^{\prime})
I_{n-1,k}(x)\\
\end{array}\right).
\end{equation}
As noted in ref. \cite{per}, one is not restricted to use equilibrium
distributions in this approach. Single particle non-equilibrium
distributions
may be more appropriate if e.g. an electric field has driven
the system
out of equlibrium.\\ \\
{\it The effective action.}$\,\,$
The generating functional for fermionic Green's functions in an external
magnetic
field may be written as a path integral:
\begin{equation}
Z(\eta,\overline{\eta},A_{\mu})=\int {\cal D}\psi\,{\cal D}
\overline{\psi}\exp
\left [\,i\int d^{\,3}x \,(-\frac{1}{4}F_{\mu\nu}F^{\mu\nu}+
\overline{\psi}(iD\!\!\!\!/ -m)\psi-\overline{\eta}\psi+\overline{\psi}\eta )
\,\right ].
\end{equation}
The functional integral describes the interaction of fermions with a
classical
electromagnetic field. It includes the effects of all virtual
electron-positron
pairs, but virtual photons are not present, which means that we are
considering
the weak coupling limit.\\ \\
The fermion field can be integrated over since the functional integral
is Gaussian:
\begin{equation}
Z(\eta,\overline{\eta},A_{\mu})=\det [\,i(iD\!\!\!\!/-m)\,]\exp
\Big [\,i\int d^{\,3}x\,\,\big [\,-\frac{1}{4}F_{\mu\nu}F^{\mu\nu}+
\int d^{\,3}y \overline{\eta}(x)S_{F}(x,y)\eta (y)\,\big ]\,\Big ].
\end{equation}
Taking the logarithm of $Z(\eta,\overline{\eta},A_{\mu})$ with vanishing
sources gives the effective action
\begin{equation}
\label{eq:seff}
S_{eff}=\int d^{\,3}x \left [\,-\frac{1}{4}F_{\mu\nu}F^{\mu\nu}\,\right ]
-i\,Tr\log
\left[\,i(iD\!\!\!\!/-m)\, \right ],
\end{equation}
where we have written $\log\det =Tr\,\log$ by the use of a complete
orthogonal
basis.  Differentiating eq. (\ref{eq:seff}) with respect to $m$ yields
\begin{equation}
\frac{\partial {\cal L}_{1}}{\partial m}=i\,trS_{F}(x,x).
\end{equation}
The trace is now over spinor indices only.
By calculating the trace of the propagator and integrating this
expression with
respect to $m$ thus yields the one-loop contribution to the effective
action.
This method has been previously applied by Elmfors {\it et al}
\cite{per} in 3+1 dimensions. The above equation may readily be
generalized to finite temperature, where we separate the vacuum
contribution in the effective action
\begin{equation}
{\cal L}={\cal L}_{0}+{\cal L}_{1}+{\cal L}^{\beta,\mu}
\equiv {\cal L}_{0}+{\cal L}_{eff}
\end{equation}
where ${\cal L}_{0}$ is the tree level contribution, and
\begin{equation}
\frac{\partial {\cal L}_{eff}}{\partial m}=i\,tr\left [ \,S_{F}(x,x)
+S_{F}^{\beta ,\mu}(x,x)\,\right ].
\end{equation}
Using eqs. (\ref{eq:fp}) and (\ref{eq:matrix}) a straightforward
calculation
gives for the vacuum contribution
\begin{eqnarray}\nonumber
trS_{F}(x,x)&=&-\frac{1}{4\pi^2}\sum_{n=1}^{\infty}\int
\frac{dk\,d\omega}{\omega^{2}-E^{2}_{n}+i\varepsilon}
\left [\, m\left (I_{n,k}^{2}(x)+I_{n-1,k}^{2}(x)\right )+
\omega\left (I_{n,k}^{2}(x)-I_{n-1,k}^{2}(x)\right )
\,\right ] \\ \nonumber
&&+ \frac{1}{4\pi^2}\int\frac{dk\,d\omega}{\omega + m - i\varepsilon}
I_{0,k}^2(x) \\ \nonumber
&=&\frac{i}{2\pi}\sum_{n=1}^{\infty}\int dk \frac{m}{E_{n}}
I_{n,k}^{2}(x) + \frac{ieB}{4\pi}\\
&=&\frac{ieB}{2\pi}\sum_{n=1}^{\infty}\frac{m}{E_{n}}+\frac{ieB}{4\pi}.
\end{eqnarray}
Integrating this expression with respect to $m$ yields
\begin{equation}
{\cal L}_{1}=-\frac{eB}{2\pi}\sum_{n=1}^{\infty}\sqrt{m^{2}+2eBn}-
\frac{eBm}{4\pi}.
\end{equation}
The divergence may be sidestepped by using the integral representation
of the
gamma function \cite{tab} and
subtract a constant to make ${\cal L}_{1}$ vanish for $B=0$,
\begin{equation}
\label{eq:lvac}
{\cal L}_{1}=\frac{1}{8\pi^{\frac{3}{2}}}\int_{0}^{\infty}
\frac{ds}{s^{\frac{5}{2}}}
\exp (-m^{2}s)\left [\, eBs\coth (eBs)-1\,\right ].
\end{equation}
This result calls for a few comments. Firstly, one notes that the
induced Chern-Simons term in the effective action vanishes. This is
not in conflict with the result of Redlich; rather it is caused by our
choice of gauge, $A_{0}=0$. Insertion of $A_{0}=0$ into the general
expression for the Chern-Simons term
obtained by Redlich, will make it vanish. Thus, our results
agree.
Secondly, one can show that the Chern-Simons term
is gauge-invariant up to surface terms and using another gauge the
Chern-Simons term needs not vanish. The choice $A_{0}\neq 0$ gives rise to
a Chern-Simons term $A_{0}Bme^{2}/(|m|8\pi^{2})$ in this approach and the
result is again in
accordance with the general expression.
It is always satisfactory to see that identical results can be obtained by
entirely different methods.\\ \\
The finite temperature part of the effective action is calculated analogously
using the thermal part of the propagator (\ref{eq:thermpro}),
\begin{eqnarray}
{\cal L}^{\beta, \mu}&=&\frac{TeB}{2\pi}\sum_{n=1}^{\infty}
\Big [\,\log [\,1+\exp -\beta(E_{n}-\mu)\,]+
\log [\,1+\exp -\beta (E_{n}+\mu )\,]\,
\Big ] \\ \nonumber
&&+\frac{TeB}{2\pi}\log \left [ \,1+\exp -\beta(m-\mu)\,\right].
\end{eqnarray}
Letting $B\rightarrow 0$ it can be shown that one obtains the free
energy of a gas of non-interacting electrons and positrons:
\begin{equation}
\label{eq:nullb}
{\cal L}_{0}^{\beta , \mu}=\frac{T}{2\pi}\int_{0}^{\infty}EdE
\Big [\,\log [\,1+\exp -\beta(E-\mu)\,]+
\log  [\,1+\exp -\beta(E+\mu)  \,]\,\Big ],
\end{equation}
where $E=\sqrt{m^{2}+k^{2}}$.\\ \\
In the following we restrict ourselves to the case $\mu >0$. Analogous
results can be obtained for $\mu <0$.  \\ \\
In the zero temperature limit of ${\cal L}^{\beta ,\mu}$ one gets
\begin{equation}
\label{eq:lo}
{\cal L}^{\beta,\mu}=\frac{eB}{2\pi}\sum_{n=0}^{\,\prime}(\mu-E_{n}),
\end{equation}
where the prime indicates that the sum is restricted to integers less than
$(\mu^{2}-m^{2})/2eB$. Similarly, one may derive the charge density at
$T=0$:
\begin{equation}
\label{eq:rho}
\rho = \frac{\partial {\cal L}^{\beta, \mu}}{\partial\mu} =
\frac{eB}{2\pi}\Big [Int(\frac{\mu^2 - m^2}{2eB})+1\Big ],\hspace{1cm}\mu>m,
\end{equation}
in accordance with the result of Zeitlin \cite{zeit}.
Notice that at $T=0$ the charge density is equal to the particle
density, since
no antiparticles are present, as can be seen by inspection of eqs.
(\ref{eq:lo}) and (\ref{eq:rho}).
\section{Magnetization and the de Haas-van Alphen Effect}
In this section we study the physical content of the effective action
which was obtained in the previous section. In particular we investigate a
few limits to check the consistency of our calculations.\\ \\
The magnetization is defined by
\begin{equation}
M=\frac{\partial {\cal L}_{eff}}{\partial B}.
\end{equation}
The vacuum contribution to the magnetization is obtained from eq.
(\ref{eq:lvac})
\begin{equation}
\label{eq:magv}
M_{1}=\frac{1}{8\pi^{\frac{3}{2}}}\int_{0}^{\infty}ds\frac{\exp \,
(-m^{2}s)}{s^{\frac{5}{2}}}\left [\,es\coth (eBs)- \frac{e^{2}Bs^{2}}
{\sinh^{2}(eBs)}\,\right ].
\end{equation}
For the thermal part of the magnetization we find
\begin{eqnarray} \nonumber
\label{eq:magt}
M^{\beta,\mu}&=&\frac{Te}{2\pi}\sum_{n=1}^{\infty}\Big  [\,
\log [\,1+\exp -\beta(E_{n}-\mu)\,]+\log [\,1+\exp -\beta(E_{n}+\mu)\,]
\,\Big] \\ \nonumber
&&+\frac{Te}{2\pi}\log \big [\, 1+\exp -\beta(m-\mu)\,\big] \\
&&-\frac{e^{2}B}{2\pi}\sum_{n=1}^{\infty}\frac{n}{E_{n}}
\Big [\, \frac{1}{\exp\beta (E_{n}-\mu)+1}+\frac{1}{\exp
\beta (E_{n}+\mu)+1}\,\Big].
\end{eqnarray}
{\it Magnetization at zero temperature.}$\,\,$
In the zero temperature limit eq. (\ref{eq:magt}) reduces to
\begin{equation}
\label{eq:st}
M^{\beta,\mu}=\frac{e}{2\pi}\sum^{\prime}_{n=0}\left[\,
\mu-E_{n}-\frac{eBn}{E_{n}} \, \right ],
\end{equation}
where the sum again is restricted to integers less than
$(\mu^{2}-m^{2})/2eB$.

In the weak $B$-field limit ($eB\ll \mu^{2}-m^{2}\ll m^{2}$)
the vacuum contribution becomes
\begin{equation}
M_{1}=\frac{e^{2}B}{12\pi^{3/2}}\int_{0}^{\infty}ds\frac{
\exp (-m^{2}s)}{s^{\frac{1}{2}}}=\frac{e^{2}B}{12\pi |m|}.
\end{equation}
This agrees with the results of refs. [14,15].
The contribution from real thermal particles is obtained by rewriting the
square root in ${\cal L}^{\beta ,\mu}$, using the integral
representation of the gamma function and
treating $Int(\mu^{2}-m^{2})/2eB$ as a continuous variable in the limit
$B\rightarrow 0$. The result is
\begin{equation}
M^{\beta,\mu}=\frac{e}{4\pi}(\mu -m).
\end{equation}
Some comments are in order. It is perhaps somewhat surprising that the
magnetization is non-zero in this limit. One should, however, bear in mind
that the sign of $m$ uniquely determines the spin of the particles
(and antiparticles), implying that the system under investigation
consists
entirely of either spin up or spin down particles. This is not the case
in 3+1d, where the representations
characterized by the sign of $m$ are equivalent. It would therefore
be natural to consider a system consisting of an equal number of spin up
and spin
down particles, which then
amounts to sum over $\pm m$~\footnote{One should
use the same chemical potential irrespective of the sign of $m$ in order
to
ensure the same charge density for spin up and spin down particles.}. By
doing so one finds a vanishing magnetization as $B$ goes to zero,
exactly as in 3+1
dimensions. Note that this result, of course, can be obtained by using
four component Dirac spinors, since the four dimensional representation
reduces to the two inequivalent two dimensional ones.\\ \\
In order to get the strong field limit $(\,eB\gg \mu^{2},m^{2})$ of the
vacuum contribution, we scale out
$eB$ and take $eB\rightarrow \infty$ in the remainder. This gives
\begin{equation}
{\cal L}_{1}\propto (eB)^{\frac{3}{2}}\Rightarrow M_{1}\propto
e^{\frac{3}{2}}\sqrt{B}.
\end{equation}
Vacuum effects contribute to the magnetization proportional to the
square root
of $B$. This should be compared
with the corresponding result in 3+1d, where the magnetization goes like
$B\log (\frac{B}{m^{2}})$ \cite{per}.
For real thermal particles only the lowest Landau level contributes in
the
strong field limit, as can be seen directly from
eq. (\ref{eq:st})~\footnote{This is also the case in 3+1d where the
energy of the lowest Landau level is independent
of B. $E_{0}=\sqrt{m^{2}+k^{2}_{z}}$, where $k_{z}$ is
the $z$-component of the
momentum.}. The thermal part of the magnetization is then
$M^{\beta,\mu} = e(\mu - m)/2\pi$. For large $B$-fields all particles are in
the lowest Landau level and $\mu = m$. Hence, the contribution to the
magnetization from real thermal particles vanishes. This can be understood
from the following physical argument:
The energy of the single particle ground
state is independent of the external field ($E_{0}=m$), so increasing
$B$ cannot
lead to an increase in ${\cal L}^{\beta,\mu}$, when the charge density
(and
therefore the particle density) is held constant.

\begin{figure}[b]
\underline{FIGURE CAPTIONS:}
\caption{The energy spectra of Dirac fermions in the presence of a
magnetic field. a) $m>0$ and b) $m<0$.}
\label{spec}
\caption{The magnetization in units of $em$ as a function of $B$ in units
of $m^{2}/e$ for different values of temperature. $\mu=1.5m$.}
\label{mag1}
\caption{A magnification of the oscillatory region in fig. 2.\hspace{5cm}}
\label{mag2}
\caption{The magnetization for a system consisting of an equal number of
spin up and spin down particles.\hspace{5cm}}
\label{mag3}
\end{figure}
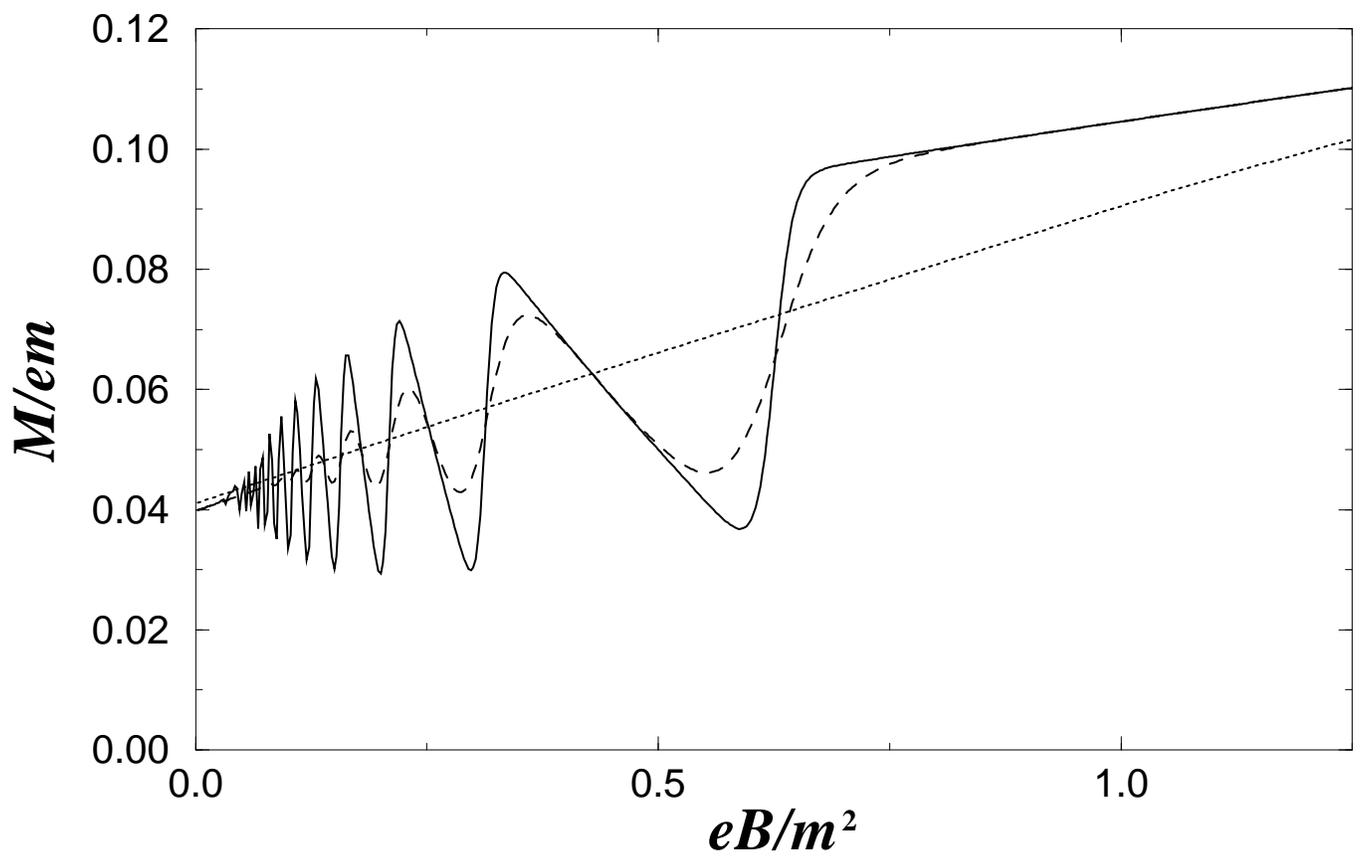
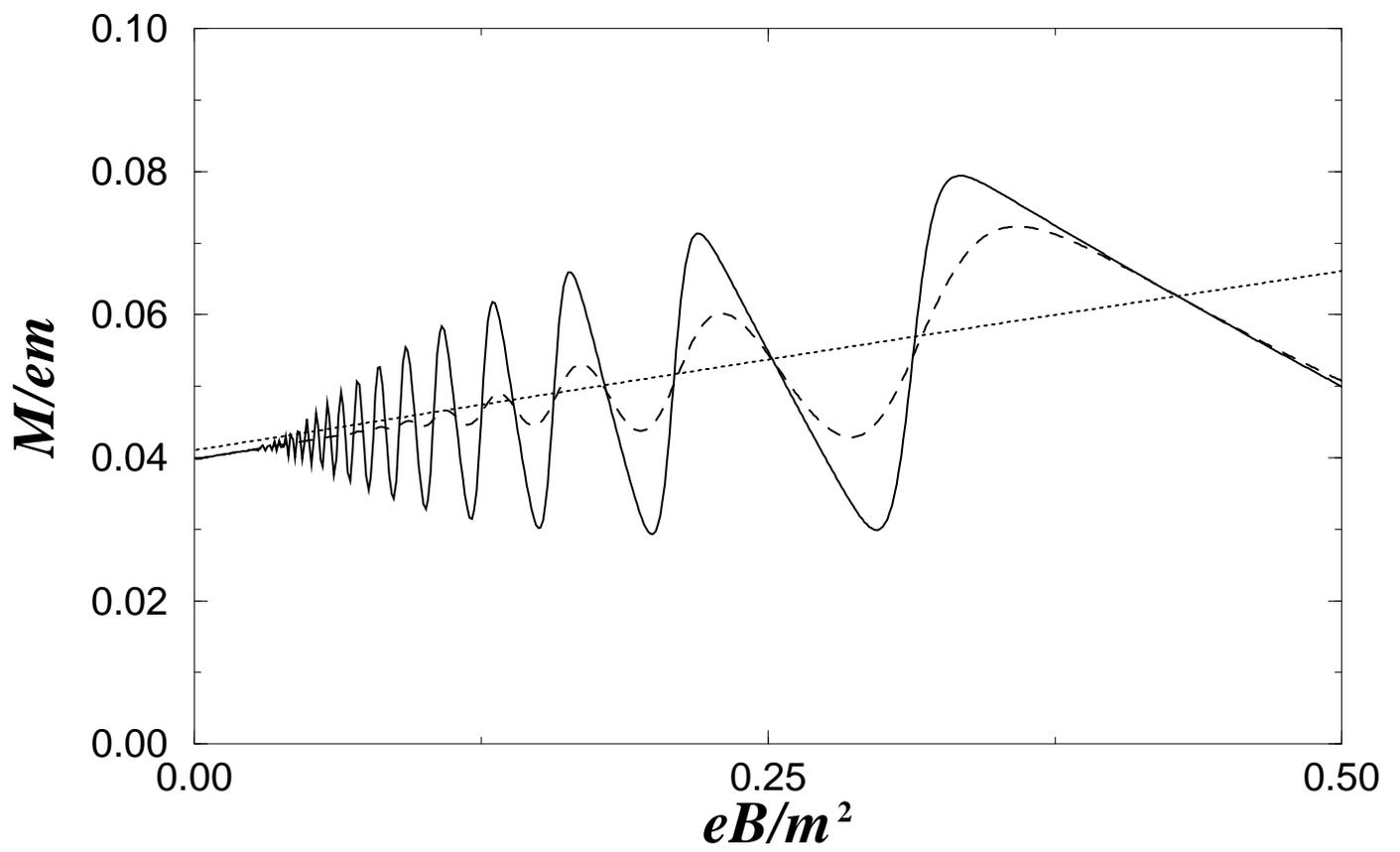
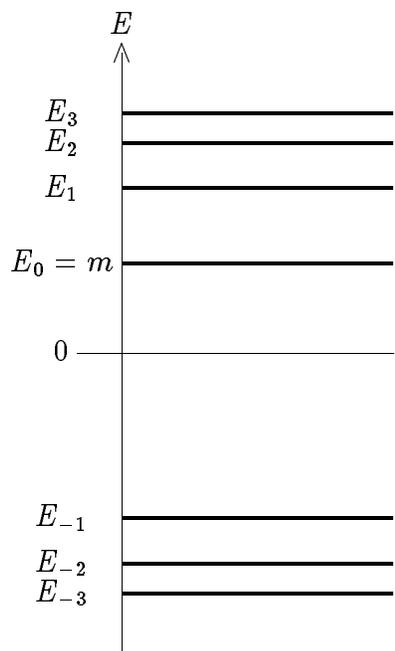
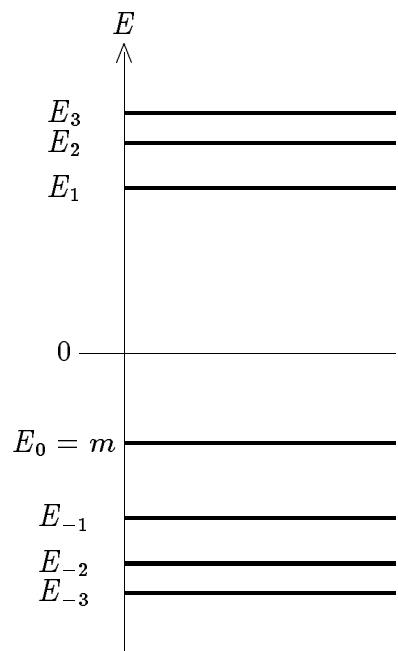
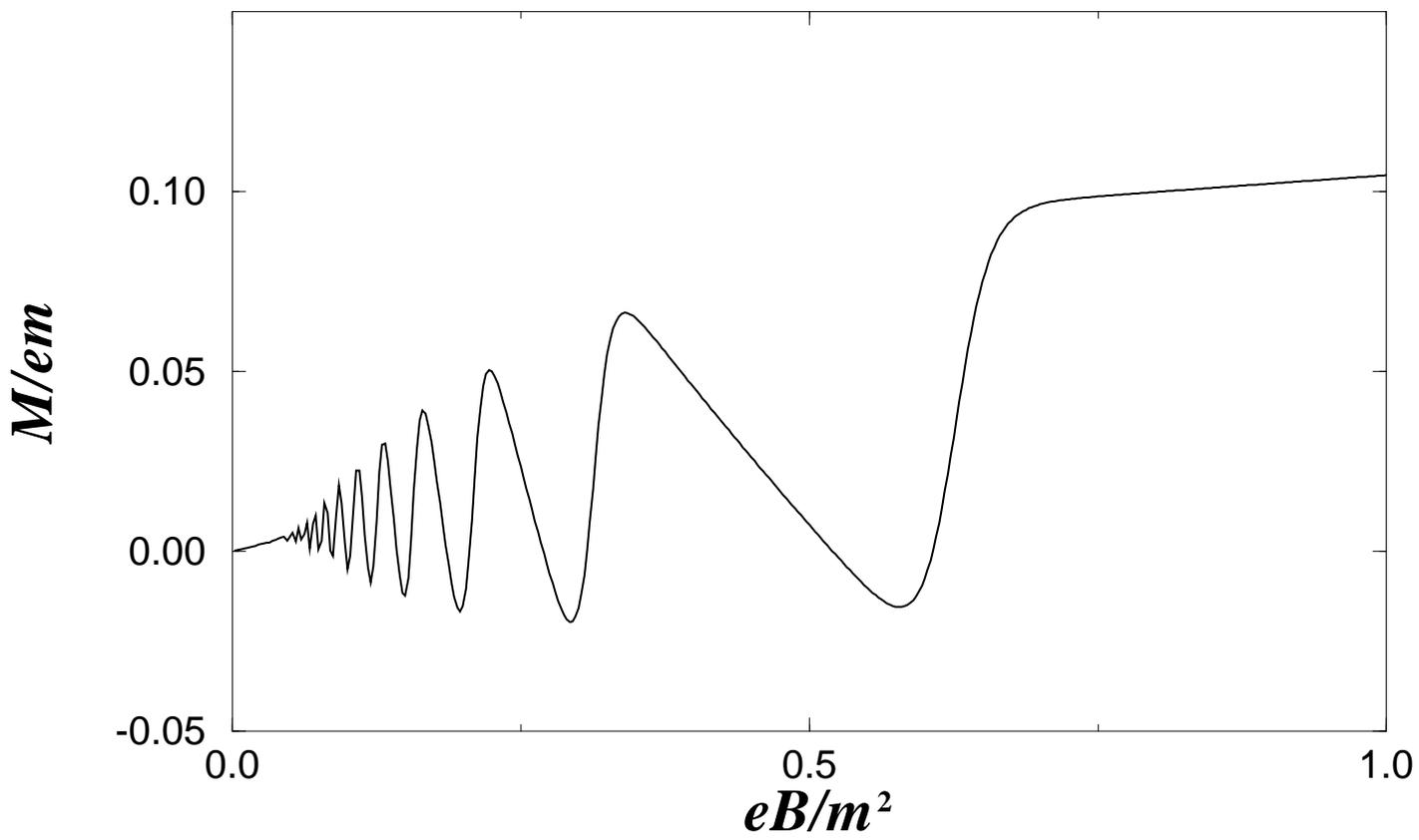

{\it High Temperature Limit $(\,T^{2}\gg m^{2}\gg eB,\mu=0)$.}$\,\,$
The high temperature limit is rather trivial. From a physical point of
view,
one expects that ${\cal L}^{\beta ,\mu}$ approaches the thermodynamic
potential of a gas of non-interacting
particles of mass $m$. Indeed we obtain
\begin{equation}
\label{eq:ht}
{\cal L}^{\beta ,\mu}=\frac{mT^{2}}{2\pi}Li_{2}[\,1 + \exp (-\beta m)\,]+
\frac{T^{3}}{2\pi}Li_{3}[\,1 + \exp (-\beta m)\,],
\end{equation}
which, of course, coincides with eq. (\ref{eq:nullb}) when
$\mu =0$. Here $Li_n(1 + x)$ is the polylogarithmic function of order $n$:
\begin{equation}
Li_n(1 + x) =\sum_{k=1}^{\infty}\frac{x^{k}}{k^{n}}.
\end{equation}
Eq. (\ref{eq:ht}) may be obtained by carefully replacing the sum by
an integral.\\ \\
{\it Magnetization at finite temperature.}$\,\,$
In fig.~\ref{mag1} we have displayed the total magnetization as a
function
of the external magnetic field for different values of the
temperature ($\mu /m=1/1.5$,\, $T/m$ =1/150 solid line,\, 1/50
dashed line,\, 1/5 dotted line).
Fig.~\ref{mag2} is a magnification
of fig.~\ref{mag1} in the oscillatory region.
The de Haas-van Alphen oscillations are seen to be present for low
temperatures and weak magnetic fields. Finally, we have summed the
magnetizations for $\pm m$.
The resulting curve ($T/m$=1/100) is shown in fig.~\ref{mag3}
and reveals that the magnetization goes to zero in the limit
$B\rightarrow 0$.   \\ \\
{\it Induced vacuum charges and currents.}$\,\,$
Finally,  we calculate the vacuum expectation value of the induced
charge and
current densities. Such calculations have been carried out in other
contexts,
e.g. in connection with magnetic flux strings (see ref.~\cite{lein}).
We shall employ the most commonly used definition of the current operator
which can be shown to measure the spectral asymmetry  relative to the
spectrum of free Dirac particles.
\begin{equation}
j^{\mu}\,(x)=\frac{e}{2}\left [\,\overline{\Psi}_{\alpha}(x),
\left ( \gamma^{\mu}\Psi (x)_{\alpha}\right) \,\right ].
\end{equation}
The induced charge density is given by
\begin{equation}
\langle\, j^{0}\,(x)\, \rangle=\langle\,\rho \,(x)\,\rangle=
-\frac{e}{2}\sum_{n}\int dk \,sign(E_{n})\mid\!\psi_{n,k}(x)\!\mid^{2}.
\end{equation}
Using the complete set of eigenmodes as given by eq. (\ref{eq:eigenm}),
a straightforward calculation yields
\begin{equation}
\label{eq:cs}
\langle\,\rho \,(x)\,\rangle =\frac{m}{\mid\! m\!\mid}\frac{e^{2}B}{4\pi}.
\end{equation}
Eq. (\ref{eq:cs}) is simply the Chern-Simons relation and our
calculations are
thus in complete agreement with the result of Fumita and Shizuya.
This  result has the following physical interpretation: As we turn the
magnetic field on, an unpaired energy level $E=m$ emerges
(in the case $m>0$).
The number of spin down negative energy electrons therefore gets reduced
relative to the free case by a factor $\frac{eB}{4\pi}$, which is
the degeneracy per unit area. This can be interpreted as the appearance of
spin up positrons and results in a positive charge density. For $m<0$ a
similar argument applies.

A similar calculation for $\langle\, {\bf j}\,(x)\, \rangle$ reveals that
the
induced current vanishes. This result should come as no surprise due to
translational symmetry of the system. A non-vanishing vacuum current
would
arise in the presence of an external electric field and is then
attributed to the drift
of the induced vacuum charge.

\section{Conclusions}
In this paper we have calculated the effective action for fermions
moving in
a plane with a constant magnetic field orthogonal to the plane. We have
derived the magnetization from the effective action and have shown
that the system exhibits the well-known  de Haas-van Alphen oscillations
at small temperatures and low values of the magnetic field.

Finally, it would be of some interest to extend the present work.
Firstly, one should examine improvements to our
results by considering the corrections to the self-energy of the electrons
in the presence of fields and a thermal heat bath.
Secondly, one could treat the highly non-trivial problem of fermions in
slowly
varying electric and magnetic fields.

The authors would like to thank M. Burgess for useful comments
and suggestions.

\newpage\pagebreak
\end{document}